\documentclass[preprint2,10.5pt]{aastex}
\begin{document}

\title{\large{\rm{A SEARCH FOR OB ASSOCIATIONS NEAR SOUTHERN LONG-PERIOD CEPHEIDS.\\ V. AQ PUP AND V620 PUP}}}

\author{D.~G. Turner\altaffilmark{1}}
\affil{Department of Astronomy and Physics, Saint Mary's University, Halifax, NS B3H 3C3, Canada}
\email{turner@ap.smu.ca}

\author{S. van den Bergh\altaffilmark{2}, P.~F. Younger}
\affil{Dominion Astrophysical Observatory, Herzberg Institute of Astrophysics, National Research Council of Canada, 5071 West Saanich Road, Victoria, BC V8X 4M6, Canada}

\author{D.~J. Majaess}
\affil{Halifax, NS, Canada}

\author{M.~H. Pedreros\altaffilmark{1}}
\affil{Departamento de F\'{i}sica, Facultad de Ciencias, Universidad de Tarapac\'{a}, Casilla 7-D, Arica, Chile}

\author{L.~N. Berdnikov\altaffilmark{3}$^,$\altaffilmark{4}}
\affil{Moscow M.~V. Lomonosov State University, Sternberg Astronomical Institute, Moscow, Russia}

\altaffiltext{1}{Visiting Astronomer, Helen Sawyer Hogg Telescope, University of Toronto}
\altaffiltext{2}{Visiting Astronomer, Cerro Tololo Inter-American Observatory, National Optical Astronomy Observatory, which is operated by the Association of Universities for Research in Astronomy (AURA), Inc., under cooperative agreement with the National Science Foundation}
\altaffiltext{3}{Visiting Astronomer, Harvard College Observatory Photographic Plate Stacks, Cambridge, MA, U.S.A.}
\altaffiltext{4}{Isaac Newton Institute of Chile, Moscow Branch, Moscow, Russia}

\begin{abstract}
A photometric {\it UBV} survey is presented for 610 stars in a region surrounding the Cepheid AQ Puppis and centered southwest of the variable, based upon photoelectric measures for 14 stars and calibrated iris photometry of photographic plates of the field for 596 stars. An analysis of reddening and distance for program stars indicates that the major dust complex in this direction is $\sim1.8$ kpc distant, producing differential extinction described by a ratio of total-to-selective extinction of $R = A_V/E_{B-V} = 3.10\pm0.20$. Zero-age main-sequence fitting for the main group of B-type stars along the line of sight yields a distance of $3.21 \pm0.19$ kpc ($V_0-M_V = 12.53 \pm 0.13$ s.e.). The 29$^{\rm d}$.97 Cepheid AQ Pup, of field reddening $E_{B-V}=0.47\pm0.07$ ($E_{B-V}{\rm (B0)}=0.51\pm0.07$), appears to be associated with B-type stars lying within $5\arcmin$ of it as well as with a sparse group of stars, designated Turner~14, centered south of it at J2000.0 = 07:58:37, --29:25:00, with a mean reddening of $E_{B-V}=0.81\pm0.01$. AQ Pup has an inferred luminosity as a cluster member of $\langle M_V \rangle = -5.40\pm0.25$ and an evolutionary age of $3\times10^7$ yr. Its observed rate of period increase of $+300.1\pm1.2$ s yr$^{-1}$ is an order of magnitude larger than what is observed for Cepheids of comparable period in the third crossing of the instability strip, and may be indicative of a high rate of mass loss or a putative fifth crossing. Another sparse cluster, designated Turner~13, surrounds the newly-recognized 2$^{\rm d}$.59 Cepheid V620 Pup, of space reddening $E_{B-V}=0.64\pm0.02$ ($E_{B-V}{\rm (B0)}=0.68\pm0.02$), distance $2.88\pm0.11$ kpc ($V_0-M_V = 12.30 \pm 0.08$ s.e.), evolutionary age $10^8$ yr, and an inferred luminosity as a likely cluster member of $\langle M_V \rangle = -2.74\pm0.11$. V620 Pup is tentatively identified as a first crosser, pending additional observations.
\end{abstract}


\section{{\rm \footnotesize INTRODUCTION}}
The most important Galactic calibrators for the Cepheid period-luminosity (PL) relation are long-period pulsators, which are less frequently found in open clusters than their short-period cousins \citep[e.g.,][]{tb02,tu10}. Such objects are massive and young enough, however, to belong to older portions of OB associations, which can often be delineated by photometric or spectroscopic methods. That philosophy initiated a program by Sidney van den Bergh thirty years ago to identify associated young B-type stars in the vicinity of bright southern hemisphere long-period Cepheids \citep*{vd82,vd83,vd85,te93}, with offshoots involving studies of potential coincidences of long-period Cepheids with open clusters \citep{te05,te09a}. The present study involves the 29$^{\rm d}$.97 Cepheid AQ Puppis ($\ell=246\degr.1562$, $b=+0\degr.1061$), which presents unique complications arising from the high degree of differential reddening by interstellar dust along its line of sight.

The rediscovery of Cepheids in Galactic open clusters by Irwin in 1955 \citep{ir55,ir58} was accompanied by additional, independent searches for Cepheid-cluster coincidences \citep{kh56,kr57,vd57,ti59}. A later study by \citet*{ts66} with an updated cluster database revealed the spatial coincidence of AQ Pup with the coronal region of the cluster Ruprecht 43, but without further follow-up, possibly because of an uncertain nature for the cluster. A possible association of AQ Pup with Pup OB2 at $d\simeq4.4$ kpc was studied by \citet{fe66}, but without definitive conclusions. \citet{gr68} took a more positive view of the same data while revising the distance of Pup OB2 to 2.9 kpc, subsequently confirmed by \citet{ts70}. An alternate possibility for an association of AQ Pup with Pup OB1 at $d\simeq2.5$ kpc was suggested by \citet{hv72}, although \citet{tu81} argued that the Cepheid appeared unlikely to be a member of either association.

Star counts in the immediate vicinity of the Cepheid did reveal a slight density enhancement \citep[Turner, see][]{eu94}, suggesting the possibility that the region near AQ Pup might contain the sparse remains of an uncatalogued open cluster or association, now designated as Turner 12 \citep{di02,al12}. A preliminary assessment \citep{tb02} designated the group as Pup OB3, for lack of a more definitive term. The present photometric survey of the field explores the preliminary findings further to reveal the possible open cluster connection that exists.

Subsequent to the original data collection and measurement an additional Cepheid was found to lie in the survey field, the $2^{\rm d}$.59 Cepheid V620 Puppis \citep[$\ell=246\degr.3115$, $b=-0\degr.1264$,][]{ka08}, originally NSV 03832 but recognized as a classical Cepheid from the ASAS-3 survey \citep{po02}. By happy circumstance V620 Pup appears to lie in a previously unrecognized sparse open cluster, so is itself a potential calibrator for the short-period end of the PL relation. The present study also discusses the independent case for its cluster membership.

There is another long-period Cepheid in the region of AQ Pup, namely the 14$^{\rm d}$.15 variable LS Pup. However, it falls just west of the surveyed region, so is not discussed here.

This is the final study in a series that was initially based on photographic photometry tied to skeleton photoelectric sequences. It has been a very large project that has by necessity extended over a number of decades. For the sake of homogeneity it was necessary to use the same techniques (e.g., iris photometry) for all program fields, although there have been improvements to the original methodology, such as incorporating additional photoelectric standards, refining the iris photometry techniques to improve the precision of the results, and completing more comprehensive analyses of interstellar reddening, particularly differential reddening, which has been ubiquitous in all survey fields.

The introduction of CCD detectors in the intervening years has changed the nature of photometric surveys. Greater precision could be achieved at present through use of a CCD detector, although mosaicing would be needed to cover the fields studied. Accurate corrections for the effects of interstellar reddening is also best achieved using {\it UBV} photometry, which can be a challenge to achieve with the panchromatic response of most CCD devices \citep[see][]{ma02}. In the end what counts is the result obtained, not the technique used, as the present study demonstrates.

\begin{figure}[!t]
\epsscale{0.90}
\plotone{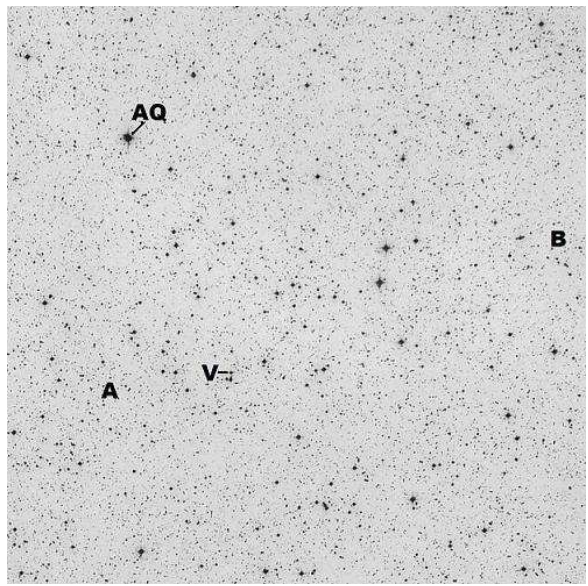}
\caption{\small{A chart (north up, east left) for the $38\arcmin$ diameter field near AQ Pup (AQ) and V620 Pup (V)  centered on J2000 = 07:57:30.7, --29:17:54.8, derived from the red ESO-SRC image. The regions designated as Fields A and B in the text are denoted by the letters A and B.}}
\label{fig1}
\end{figure}

\section{{\rm \footnotesize OBSERVATIONAL DATA}}
The data for the present study include photoelectric {\it UBV} photometry of 14 stars in the field of AQ Pup from observations obtained with the Cerro Tololo 1.5-m telescope in March 1979 and the University of Toronto's Helen Sawyer Hogg 0.6-m telescope in March 1976 and January 1987, when it was located on Cerro Las Campanas, Chile. Details are provided by \citet{tu81}, \citet{vd82}, \citet*{sh04}, and \citet{te09a}. The data for the 14 stars used as photoelectric standards are summarized in Table~\ref{tab1}, where the stars are identified by Arabic letters and their co-ordinates in the 2MASS catalogue \citep{cu03}. Three of the stars are numbered in the {\it Catalogue of Luminous Stars in the Southern Milky Way} \citep{ss71}.

\begin{deluxetable}{@{\extracolsep{-2.5mm}}ccccccc}
\tabletypesize{\small}
\tablewidth{0pt}
\tablenum{1}
\tablecaption{Photoelectric data for the AQ Pup field
\label{tab1}}
\tablehead{
\colhead{Star} &\colhead{RA(2000)} &\colhead{Dec(2000)} &\colhead{{\it V}} &\colhead{{\it B--V}}  &\colhead{{\it U--B}} &\colhead{n} } 
\startdata
AQa &07:58:22.442 &--29:07:39.57 &11.13 &1.33 &+1.45 &10 \\
AQb &07:58:23.054 &--29:07:37.61 &12.74 &0.21 &+0.14 &10 \\
A &07:58:24.378 &--29:07:52.60 &14.70 &0.60 &+0.26 &1 \\
B &07:58:09.256 &--29:09:47.08 &10.23 &0.36 &+0.13 &3 \\
C &07:58:20.117 &--29:10:15.86 &12.50 &0.53 &--0.29 &3 \\
D &07:58:20.818 &--29:09:53.60 &14.45 &0.63 &+0.18 &1 \\
E &07:58:22.950 &--29:10:19.76 &13.49 &0.50 &+0.28 &3 \\
F &07:58:19.870 &--29:11:15.82 &11.58 &0.49 &--0.34 &3 \\
G &07:58:18.654 &--29:11:59.54 &13.01 &0.24 &+0.19 &3 \\
H &07:58:21.097 &--29:05:57.74 &13.31 &0.62 &+0.13 &3 \\
I &07:58:24.257 &--29:06:36.08 &14.90 &0.53 &+0.41 &1 \\
J\tablenotemark{a} &07:57:56.655 &--29:05:47.47 &11.63 &0.19 &--0.59 &3 \\
K\tablenotemark{b} &07:57:25.109 &--29:10:03.75 &9.79 &0.20 &--0.70 &8 \\
L\tablenotemark{c} &07:56:26.407 &--29:25:26.13 &10.28 &0.43 &--0.44 &5 \\
\enddata
\tablenotetext{a}{~LSS 890}
\tablenotetext{b}{~LSS 888}
\tablenotetext{c}{~LSS 876}
\end{deluxetable}

The photoelectric data were supplemented by photographic {\it UBV} photometry for 596 stars in the vicinity of AQ Pup obtained from iris photometry of plates in {\it U}, {\it B}, and {\it V} taken with the 0.9-m Swope telescope on Cerro Las Campanas in May 1978 and February 1982 (2 plates in {\it B} and 1 plate in each of {\it V} and {\it U}). The iris measures were obtained using the Cuffey Iris Astrophotometer at Saint Mary's University along with the prescriptive techniques described by \citet{tw89}, which are designed to generate data with a precision approaching $\pm0.03$ magnitude. The resulting photographic measures encompass a field of $\sim18\arcmin$ radius centered on J2000.0 coordinates 07:57:30.696, --29:17:54.79 (Fig.~\ref{fig1}), and are summarized in Table~\ref{tab2}.

The effectiveness of surveys such as this varies in direct proportion to the accuracy and precision of the input data. The precision is limited by photon counting statistics for the photoelectric observations and by photographic grain noise for the photographic photometry, and in our experience is typically $\pm0.01$ and $\pm0.03$ to $\pm0.04$, respectively, in both magnitudes and colors. The techniques adopted by \citet{tw89} have been demonstrated to reach such a level of precision for iris measures of photographic plates, provided that steps are taken to measure complete stellar images, including extended tails, and which lead to magnitude calibrations that are simple power laws of the iris readings. That was the case here. The accuracy of the observations is generally tested by comparison with the results of other studies for stars in common and by direct examination of the data, as illustrated here.

\begin{deluxetable}{@{\extracolsep{-2.5mm}}ccccrl}
\tabletypesize{\small}
\tablewidth{0pt}
\tablenum{2}
\tablecaption{Photographic data for the AQ Pup field
\label{tab2}}
\tablehead{
\colhead{RA(2000)} &\colhead{Dec(2000)} &\colhead{{\it V}} &\colhead{{\it B--V}} &\colhead{{\it U--B}}  &\colhead{Comments} }
\startdata
07:56:06.216 &--29:14:20.10 &12.78 &0.67 &--0.46 & \\
07:56:06.556 &--29:17:36.63 &12.88 &1.67 &1.17 & \\
07:56:07.208 &--29:21:34.57 &12.29 &0.78 &0.07 & \\
07:56:07.554 &--29:20:15.90 &13.55 &0.49 &--0.10 & \\
07:56:08.866 &--29:16:05.84 &11.97 &0.71 &--0.13 & \\
07:56:09.486 &--29:21:04.73 &13.17 &0.44 &0.38 & \\
07:56:09.679 &--29:15:34.22 &13.29 &0.66 &0.60 & \\
07:56:10.370 &--29:15:27.01 &12.30 &1.03 &0.65 & \\
07:56:12.414 &--29:15:20.09 &12.71 &0.55 &--0.07 & \\
07:56:12.870 &--29:15:05.31 &12.81 &0.56 &--0.26 & \\
\enddata
\tablecomments{Table~2 is published in its entirety in the electronic version of the Astronomical Journal. A portion is shown here for guidance regarding its form and content.}
\end{deluxetable}

\begin{figure}[!t]
\epsscale{0.90}
\plotone{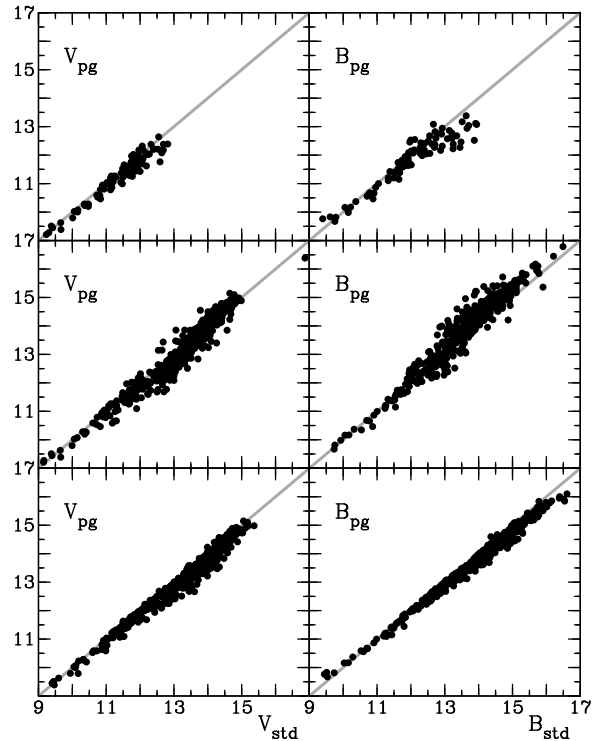}
\caption{\small{A comparison of the present photographic {\it V} (left) and {\it B} (right) magnitudes with those measured by \citet[][top]{esa97}, NOMAD \citep[][middle]{za05}, and the AAVSO's APASS program (bottom). Gray lines denote the correlation expected for perfect agreement.}}
\label{fig2}
\end{figure}

There is no other source of {\it UBV} photometry available for the general field of AQ Pup, but recent surveys have generated {\it BV} photometry for most program stars. A comparison of our photometry (Fig.~\ref{fig2}) with that of \citet{esa97} from the {\it Hipparcos and Tycho Catalogues} reveals no obvious discrepancies. The photographic {\it V} magnitudes agree closely with the \citet{esa97} results for stars brighter than $V=11$ and stars of $11<V<12$, although the scatter is larger for the latter, which are near the limit for ESA photometry. The photographic {\it B} magnitudes also agree closely with the \citet{esa97} results for stars brighter than $B=12$ and stars of $12<B<12.6$, for which the scatter is larger. There are systematic trends in both {\it V} and {\it B} for fainter stars, which are probably beyond the true limits for ESA photometry. A similar result was found in a comparison of the photographic photometry with that of \citet{kr01}, which represents a tweaking of the \citet{esa97} photometry.

The NOMAD survey resulting from calibrated scans of the POSS \citep{za05} is similar in some respects to our photographic survey, but reveals differences for stars fainter than $V=B=11$ (Fig.~\ref{fig2}) that are in opposite senses. Such deviations may reflect differences in the standard stars adopted for calibration purposes.

The APASS survey of the American Association of Variable Star Observers (AAVSO) contains CCD {\it BV} photometry for most stars in our field. A comparison (Fig.~\ref{fig2}) of {\it V} magnitudes for stars in common shows good agreement to $V\simeq15$, but there are trends for many stars fainter than $V=12$, which appear to be measured systematically fainter by APASS. A comparison of the {\it B} magnitudes is similar. There is generally good agreement for stars brighter than $B=12$, but systematic trends for most stars fainter than that, again with the stars being measured fainter by APASS. Both trends are the same as those seen in \citet{esa97} photometry for stars at the photometric limits of that survey, which suggests potential problems with the calibration of the APASS results. Of course, there is also a potential for slight non-linearity in the faint star calibration of the iris measures, but it is not clear how serious that may be.

\begin{figure}[!t]
\epsscale{0.90}
\plotone{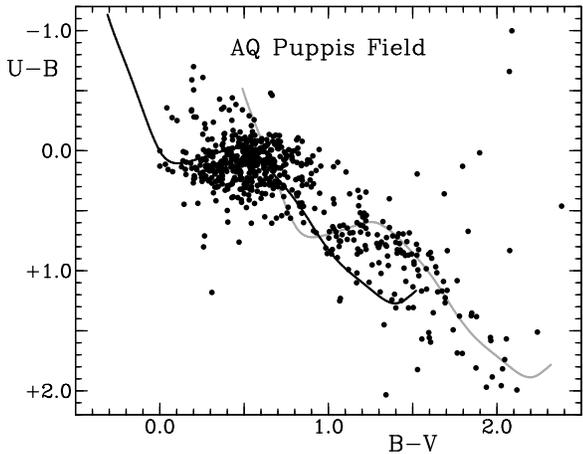}
\caption{\small{A {\it UBV} color-color diagram for measured stars in the AQ Pup field. The intrinsic relation for main-sequence stars is plotted as a solid black curve. The same relation reddened by $E_{B-V}=0.8$ is depicted by a gray curve.}}
\label{fig3}
\end{figure}

Alternatively, the data themselves provide a good impression of their overall accuracy. Fig.~\ref{fig3} is a {\it UBV} color-color diagram for the complete sample of measured stars. Evident here is a selection of perhaps 10 stars that have questionable colors, some of which are stars in Table~\ref{tab2} for which nearby companions have affected the photometry. Others may lie near the survey limits, where the photometric calibrations are uncertain. But the majority of stars in Fig.~\ref{fig2} display the colors expected for a large sample of stars affected by significant amounts of differential reddening. The bluest stars in the sample tend to be reddened stars with intrinsic colors of ({\it B--V})$_0=-0.25$, corresponding to main-sequence spectral type B2~V, implying that the Puppis OB associations contain very few members in this section of the constellation. What is striking is the selection of $\sim50$ stars that fall almost exactly on the intrinsic relation for unreddened main-sequence stars, and a similar group that lies extremely close to the intrinsic relation for AFGK-dwarfs reddened by $E_{B-V}=0.8$. If the data were affected by systematic magnitude-dependent errors, such features would simply not occur. Of course, there is scatter in the observations, but not more than expected for the cited precision.

\begin{figure}[!t]
\epsscale{0.90}
\plotone{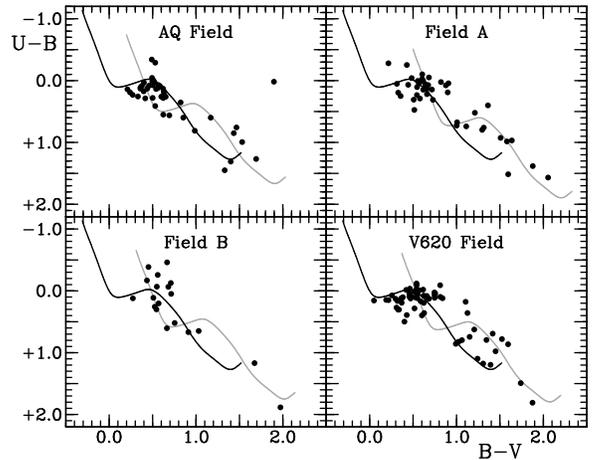}
\caption{\small{{\it UBV} color-color diagrams for separate $5\arcmin$ radius subfields in the AQ Pup region: surrounding AQ Pup (upper left), sparse cluster A centered at J2000.0 = 07:58:37, --29:25:00 (upper right), a group of stars (Field B) of common reddening centered at J2000.0 = 07:56:06, --29:13:27 (lower left), and the environs of V620 Pup (lower right). The plotted relations are those of Fig.\ref{fig3}, except the color excesses are $E_{B-V}=0.51$, 0.81, 0.62, and 0.68, respectively.}}
\label{fig4}
\end{figure}

\begin{figure}[!t]
\epsscale{0.90}
\plotone{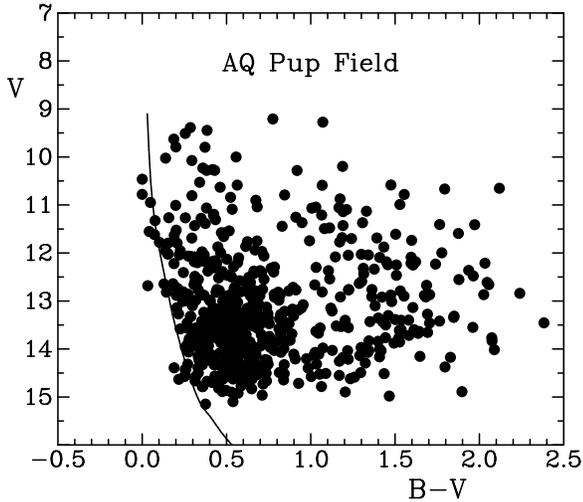}
\caption{\small{A {\it BV} color-magnitude diagram for measured stars in the AQ Pup field. The zero-age main sequence (ZAMS, black curve) is plotted for a reddening of $E_{B-V}=0.35$ and $V-M_V=13.6$ ($V_0-M_V=12.55$).}}
\label{fig5}
\end{figure}

Further evidence regarding the general accuracy of the photometric observations is provided by isolating stars according to location in the field. Fig.~\ref{fig4} contains individual color-color diagrams for four separate regions, of $5\arcmin$ radius, in the survey area: AQ Pup (J2000.0 = 07:58:22, --29:07:48), Field A (J2000.0 = 07:58:37, --29:25:00), Field B (J2000.0 = 07:56:06, --29:13:27), and V620 Pup (J2000.0 = 07:57:50, --29:23:03). Although differential reddening is present here as well, it is possible to identify small groups of stars of identical reddening, or even of little to no reddening. That also occurs in space reddening plots for the sample, where adjacent stars share similar reddenings, or zero reddening, to within $\pm0.01$ to $\pm0.02$ in $E_{B-V}$. Such results are only possible with photometry of reasonably high precision and accuracy. It implies good results for the derived space reddenings of the two Cepheids AQ Pup and V620 Pup, provided that differential reddening in their fields is not severe. The high degree of differential reddening in the field is also evident from the uncorrected color-magnitude diagram for program stars plotted in Fig.~\ref{fig5}. The scatter here is typical of fields where the color excesses $E_{B-V}$ for group stars exhibit a spread of a magnitude or more \citep{tu76b}.

\section{{\rm \footnotesize ANALYSIS}}
\begin{figure}[!t]
\epsscale{0.90}
\plotone{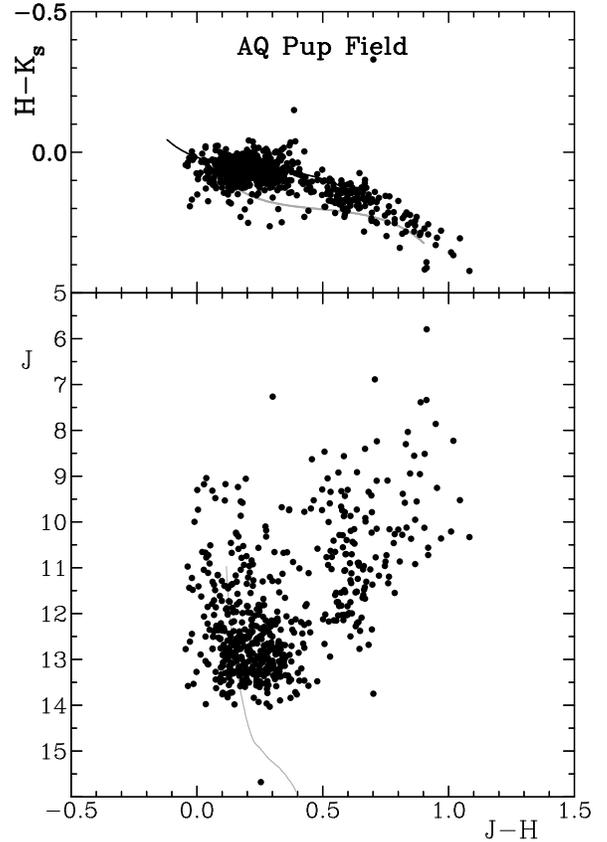}
\caption{\small{A 2MASS color-color diagram (upper) and color-magnitude diagram (lower) for measured stars in the AQ Pup field. The intrinsic relation is plotted as a black curve, while a gray curve represents the same relation for a reddening equivalent to that used in Fig.\ref{fig3}. The plotted ZAMS relation (gray curve, lower) has the same reddening for the intrinsic distance modulus of Fig.\ref{fig7}.}}
\label{fig6}
\end{figure}

\begin{figure}[!t]
\epsscale{0.90}
\plotone{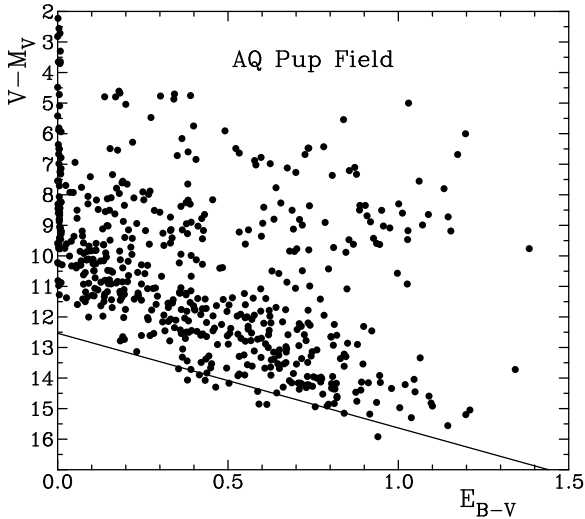}
\caption{\small{A variable-extinction diagram (plot of apparent distance modulus vs. color excess) for measured stars in the AQ Pup field, with absolute magnitudes assuming ZAMS luminosities. The line fitted to the lower envelope of the data has slope $R=3.10\pm0.20$ and zero-point $V_0-M_V=12.53\pm0.13$.}}
\label{fig7}
\end{figure}

The {\it UBV} data of Tables \ref{tab1} and \ref{tab2} plotted in Fig.~\ref{fig3} were corrected for reddening using a reddening law for the field found previously \citep{tu81,tu89}. It is described by $E_{U-B}/E_{B-V}=0.75 + 0.02\;E_{B-V}$. {\it JHK}$_s$ observations for the same stars from the 2MASS catalogue \citep{cu03} served as a guide for resolving ambiguities in likely intrinsic color for some stars, although excessive scatter in the {\it JHK}$_s$ data (Fig.~\ref{fig6}) actually introduced ambiguities of their own in many cases. The derived color excesses, $E_{B-V}$, were also normalized to those appropriate for a B0 star observed through the same amount of dust \citep{fe63}. Absolute magnitudes appropriate for zero-age main sequence (ZAMS) stars of the same intrinsic color \citep{tu76a,tu79} were adopted in order to provide data suitable for a variable-extinction analysis of the stars \citep[see][]{tu76a,tu76b}. The results are presented in Fig.~\ref{fig7}.


A small proportion of stars in the AQ Pup field are unreddened late-type stars; none are O-type stars. However, there are many B-type, A-type, and F-type stars of various reddenings present throughout the region. Unreddened stars in the sample can be detected to intrinsic distance moduli reaching values as large as $V_0-M_V=11.25$, corresponding to distances of 1.78 kpc. That is consistent with what was found previously by \citet{hv72} and \citet{nk80} for distances to the dust clouds in this direction. It appears that the Galactic plane along the line of sight is relatively dust-free for about 1.8 kiloparsecs, beyond which varied and occasionally substantial extinction arises.

The small density enhancement of stars denoted as Turner~12 at the center of the survey field of Fig.~\ref{fig1} was originally detected by star counts using the low plate scale {\it Vehrenberg Atlas} \citep{ve64}, which is superior to the POSS for detecting extended clusters. Most of the stars in that region have small color excesses relative to stars lying in the outer portions, and there appears to be a dust ring surrounding them on the north side that contains no program stars at all. The small density enhancement referred to as Turner~12 is therefore primarily an optical effect arising from patchy extinction in the field, and there is no clear photometric evidence for its existence as an extended star cluster or association.

The general trends in the data of Fig.~\ref{fig7} appear to follow a value for the ratio of total-to-selective extinction, $R = A_V/E_{B-V}$, that is close to 3. Yet the distribution of data suggests that the reddening out to 1.8 kpc may also be patchy. Note, for example, the sequences of points that clump towards intrinsic distance moduli of about 5 and 8, i.e., 100 pc and 400 pc. If the stars are post main-sequence objects rather than dwarfs, then they must have luminosities of bright giants or supergiants if reddened by extinction arising in the main dust complex 1.8 kpc distant.

There is a fairly distinct lower envelope in Fig.~\ref{fig7} of reddened stars, presumably ZAMS stars, that have reddenings typically in excess of $E_{B-V}=0.35$. Least squares and non-parametric techniques applied to that group yield best-fitting values of $R = A_V/E_{B-V} = 3.10 \pm 0.20$ s.e. and $V_0-M_V = 12.53 \pm 0.13$ s.e., corresponding to a distance of $3.21 \pm 0.19$ kpc. The value of {\it R} is consistent with expectations for dust producing a reddening slope of $E_{U-B}/E_{B-V} \simeq 0.75$ \citep{tu94,tu96}, even though the dust is not local.

\begin{figure}[!t]
\epsscale{0.90}
\plotone{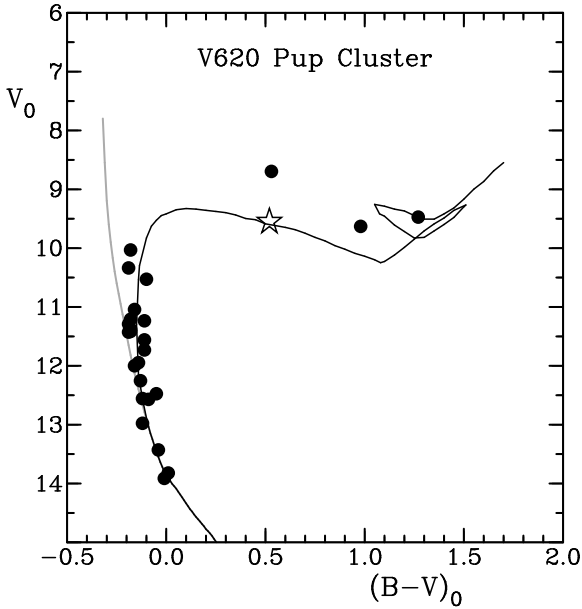}
\caption{\small{A reddening-corrected color-magnitude diagram for stars in the putative cluster surrounding V620 Pup. A gray curve represents the ZAMS for $V_0-M_V=12.30\pm0.08$, a black curve an isochrone for $\log t=8.0$, and the star symbol represents the mean parameters of V620 Pup.}}
\label{fig8}
\end{figure}

Unreddened parameters for the group of stars surrounding V620 Pup are depicted in Fig.~\ref{fig8} for an intrinsic distance modulus of $V_0-M_V=12.30\pm0.08$ derived for 6 stars that appear to lie on the ZAMS. There is a reasonably good fit of the data to a model isochrone for $\log t = 8.0$ taken from \citet{me93}. The implied distance of $2.88\pm0.11$ kpc requires confirmation from a deeper survey, and might be  too small. The reality of the cluster, designated here as Turner~13, also needs to be confirmed by star counts and radial velocities. The field immediately surrounding V620 Pup does appear to contain an above average number of faint, reddened, blue stars, despite the relatively high reddening for the cluster, and that and the data of Fig.~\ref{fig8} are presently the only evidence for the cluster's existence. Two of the red stars near the Cepheid may be red giant (GK-type) members of the cluster. They fit the $\log t=8.0$ isochrone reasonably well. The mean reddening of stars lying close to V620 Pup is $E_{B-V}{\rm (B0)}=0.68\pm0.02$ s.e., which corresponds to a field reddening for the Cepheid of $E_{B-V}=0.64\pm0.02$. The corresponding evolutionary age of Turner~13 and V620 Pup is $\sim10^8$ yr.

\begin{figure}[!t]
\epsscale{0.90}
\plotone{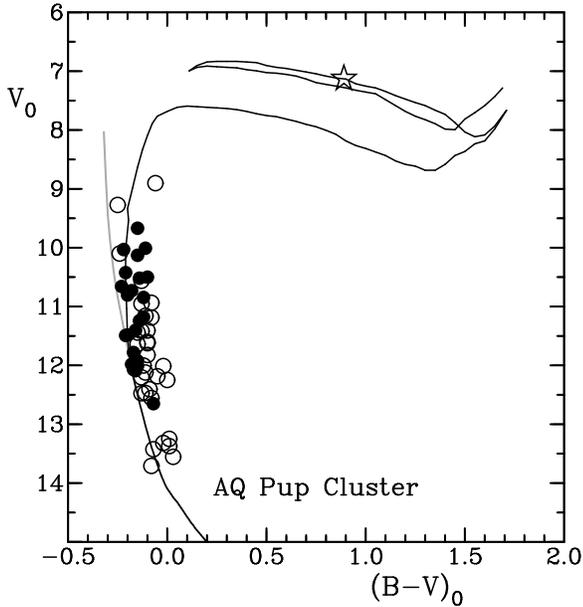}
\caption{\small{A reddening-corrected color-magnitude diagram for stars in the putative cluster south of AQ Pup (filled circles) and for stars surrounding the Cepheid (open circles). A gray curve represents the ZAMS for $V_0-M_V=12.53\pm0.13$, a black curve an isochrone for $\log t=7.5$, and the star symbol represents the mean parameters of AQ Pup.}}
\label{fig9}
\end{figure}

Unreddened parameters for the group of stars labeled as Field A in Fig.~\ref{fig3} are depicted in Fig.~\ref{fig9} for the distance modulus derived in Fig.~\ref{fig7}. Although differential reddening is very clearly present in the region of this group of stars, designated here as Turner~14, the mean reddening of stars near the cluster core is $E_{B-V}{\rm (B0)}=0.81\pm0.01$ s.e. The reality of the cluster is indicated by an increase in the density of faint blue stars in the core regions of Turner~14, despite a rather large reddening. The identification of possible cluster members on the basis solely of reddening and location in the color-magnitude diagram of Fig.~\ref{fig9} is difficult. The large apparent redward spread in the evolved portion of the cluster main sequence is an effect seen in other intermediate-age clusters \citep*[see][]{te92,tu93}, and may have a similar explanation in terms of circumstellar reddening for stars of large rotational velocity.

Additional members of Turner~14 can be found in the region surrounding AQ Pup, as shown in Fig.~\ref{fig9}. The identification of several of the stars as potential cluster members is motivated by the possibility that rapid rotation is inherent to many stars on the cluster main sequence. If that is incorrect, then 6--7 of the stars could be removed from the sample as likely foreground dwarfs. As noted in Fig.~\ref{fig4}, the mean reddening of stars lying within $2\arcmin$ of AQ Pup is $E_{B-V}{\rm (B0)}=0.51\pm0.07$ s.e., equivalent to a space reddening for the Cepheid of $E_{B-V}=0.47\pm0.07$. Differential reddening is quite strong near the Cepheid and accounts for the large uncertainty in the results. The data for possible cluster members identified in Fig.~\ref{fig9} provide a reasonably good fit to a model isochrone for $\log t = 7.5$ taken from \citet{me93}. The corresponding age of the cluster is $\sim3\times10^7$ years.

\section{{\rm \footnotesize AQ PUPPIS AND V620 PUPPIS}}
In a simulation of period changes for long-period Cepheids tied to stellar evolutionary models, \citet{pi03} argues, from only 5 observed times of light maximum, that the period of AQ Pup does not appear to be changing. Such a conclusion emphasizes the importance of studying Cepheid period changes observationally using lengthy and rich temporal samples of light curve data.

\begin{figure}[!t]
\epsscale{0.90}
\plotone{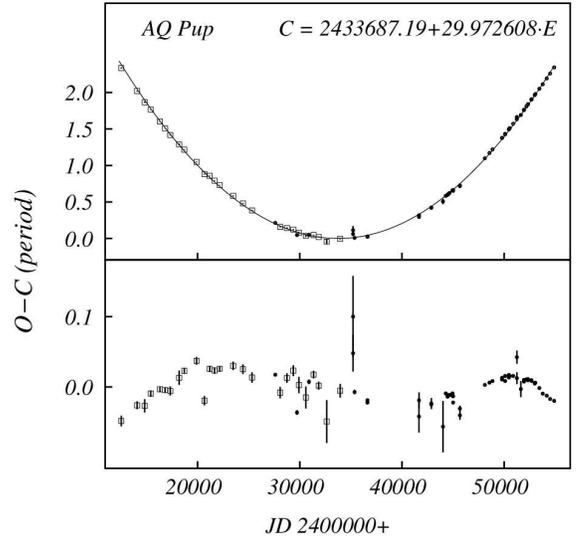}
\caption{\small{The differences between observed (O) and computed (C) times of light maximum for AQ Pup, computed in units of pulsation phase. Data based on photoelectric observations are denoted as filled circles, others as open circles. The upper diagram shows the actual O--C variations with their uncertainties, the lower diagram the residuals from the calculated parabolic evolutionary trend.}}
\label{fig10}
\end{figure}

Period changes for AQ Pup were established here from examination of archival photographic plates of the variable in the collection of the Harvard College Observatory, as well as from an analysis of new and existing photometry for the star. A working ephemeris for AQ Pup based upon the available data was:
\begin{displaymath}
{\rm JD}_{\rm max} = 2435156.13 + 29.985704 \: E ,
\end{displaymath}
where $E$ is the number of elapsed cycles. An extensive analysis of all available observations produced the data summarized in Table~\ref{tab3}, which lists results for 79 different epochs of light maximum derived from the complete light curves using Hertzsprung's method \citep{be92}, the type of data analyzed (PG = photographic, VIS = visual telescopic observations, B = photoelectric {\it B}, and V = photoelectric {\it V}), the number of observations used to establish the observed light maxima, and the source of the observations, in addition to the temporal parameters. The data are plotted in Fig.~\ref{fig10}.

A regression analysis of the O--C data of Table~\ref{tab3} produced a parabolic solution for the ephemeris defined by:
\begin{displaymath}
{\rm JD}_{\rm max} = 2433687.1920\;(\pm0.1397)
\end{displaymath}
\begin{displaymath}
+ 29.9726\;(\pm0.0002) \: E\; + 142.49\;(\pm0.56) \times 10^{-6} \:E^2 ,
\end{displaymath}
which overlays the data in Fig.~\ref{fig10}. The parabolic trend corresponds to a rapid period increase of $+300.05 \pm 1.18$ s yr$^{-1}$, a value representative of the most rapid rate of period increase for Cepheids of comparable pulsation period.

\begin{figure}[!t]
\epsscale{0.90}
\plotone{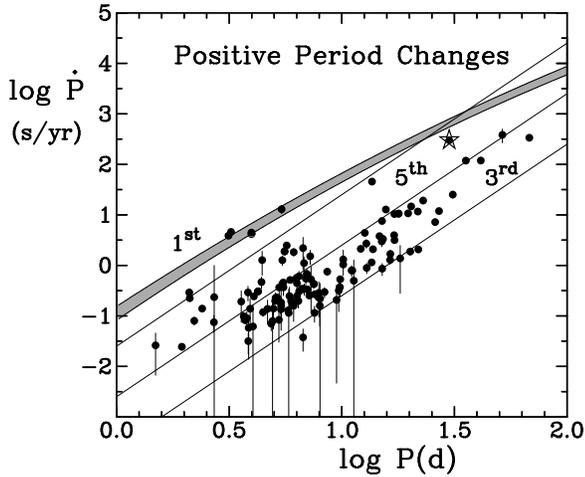}
\caption{\small{Observed and predicted rates of period change for Cepheids exhibiting period increases, with lines used to denote stars with period increases consistent with a third crossing of the instability strip, a separate group designated as putative fifth crossers, and a gray band representing predictions for first-crossing Cepheids. The rate for AQ Pup is denoted by a star symbol.}}
\label{fig11}
\end{figure}

In their study of Cepheid period changes, \citet*{te06} noted that the observed rates of period change were consistent with predictions from evolutionary models for stars in the first, second (period decreases), and third crossing of the instability strip, despite a wide range in predicted rates. The observed rates for the luminous, long-period variables also exhibited a smaller dispersion than that derived from various published models, and it was speculated that the Cepheid sample was dominated by stars in the second and third crossings of the strip, which are the slowest. Some older models, such as those of \citet{ib67} for example, predicted fourth and fifth crossings for stars in the thick helium-burning shell phase, which occurs at a more rapid rate than second and third crossings for stars of the same mass. \citet{te06} therefore speculated that the main sample of Cepheids consisted of second and third crossers, with a smaller group of putative fourth and fifth crossers exhibiting period changes at rates an order of magnitude larger than those of other Cepheids of similar period. As indicated in Fig.~\ref{fig11}, AQ Pup falls in the latter category according to its observed rate of period increase. While the location of AQ Pup in Fig.~\ref{fig9} relative to the model isochrone for $\log t = 7.5$ is consistent with either a third or fifth crossing, further evidence would be useful.

\begin{deluxetable}{@{\extracolsep{-2.5mm}}ccccccl}
\tabletypesize{\small}
\tablewidth{0pt}
\tablenum{3}
\tablecaption{Times of Maximum Light for AQ Pup
\label{tab3}}
\tablehead{
\colhead{HJD$_{\rm max}$} &\colhead{$\pm \sigma$} &\colhead{Band} &\colhead{Epoch} &\colhead{O--C}  &\colhead{Observations} &\colhead{Reference} \\
& & &\colhead{(E)} &\colhead{(days)} &\colhead{(n)} }
\startdata
2412536.650 &0.228 &PG &--708 &70.065 &9 &This paper (Harvard) \\
2414085.787 &0.168 &PG &--656 &60.627 &9 &This paper (Harvard) \\
2414830.494 &0.286 &PG &--631 &56.018 &9 &This paper (Harvard) \\
2415426.928 &0.134 &PG &--611 &53.000 &32 &This paper (Harvard) \\
2416321.197 &0.115 &PG &--581 &48.091 &50 &This paper (Harvard) \\
2416827.920 &0.107 &PG &--564 &45.279 &31 &This paper (Harvard) \\
2417334.729 &0.190 &PG &--547 &42.554 &28 &This paper (Harvard) \\
2418200.094 &0.306 &PG &--518 &38.714 &38 &This paper (Harvard) \\
2418707.457 &0.128 &PG &--501 &36.542 &27 &This paper (Harvard) \\
2419901.302 &0.165 &PG &--461 &31.483 &28 &This paper (Harvard) \\
2420675.573 &0.198 &PG &--435 &26.466 &26 &This paper (Harvard) \\
2421184.394 &0.134 &PG &--418 &25.753 &49 &This paper (Harvard) \\
2421662.015 &0.135 &PG &--402 &23.812 &30 &This paper (Harvard) \\
2422169.723 &0.123 &PG &--385 &21.986 &30 &This paper (Harvard) \\
2423484.089 &0.181 &PG &--341 &17.556 &30 &This paper (Harvard) \\
2424440.107 &0.203 &PG &--309 &14.451 &24 &This paper (Harvard) \\
2425336.419 &0.224 &PG &--279 &11.585 &18 &This paper (Harvard) \\
2427609.234 &0.047 &PG &--203 &6.482 &30 &\citet{ol36} \\
2428087.145 &0.244 &PG &--187 &4.831 &32 &This paper (Harvard) \\
2428746.059 &0.204 &PG &--165 &4.348 &32 &This paper (Harvard) \\
2429374.870 &0.231 &PG &--144 &3.733 &26 &This paper (Harvard) \\
2429732.290 &0.102 &PG &--132 &1.482 &30 &\citet{ch65} \\
2429913.074 &0.346 &PG &--126 &2.431 &34 &This paper (Harvard) \\
2430601.166 &0.470 &PG &--103 &1.153 &39 &This paper (Harvard) \\
2430901.277 &0.098 &PG &--93 &1.538 &16 &\citet{er65} \\
2431350.810 &0.163 &PG &--78 &1.481 &33 &This paper (Harvard) \\
2431859.537 &0.173 &PG &--61 &0.674 &26 &This paper (Harvard) \\
2432636.941 &0.913 &PG &--35 &--1.209 &27 &This paper (Harvard) \\
2433956.885 &0.289 &PG &+9 &--0.061 &35 &This paper (Harvard) \\
2435217.689 &0.780 &B &+51 &1.894 &8 &\citet{ir61} \\
2435219.162 &1.733 &V &+51 &3.367 &8 &\citet{ir61} \\
2435365.976 &0.094 &PG &+56 &0.318 &17 &\citet{er65} \\
2436625.298 &0.035 &B &+98 &0.791 &14 &\citet{fe66} \\
2436625.308 &0.027 &V &+98 &0.800 &16 &\citet{fe66} \\
2441668.729 &0.685 &V &+266 &8.823 &9 &\citet{ma75} \\
2441669.510 &0.334 &B &+266 &9.604 &9 &\citet{ma75} \\
2442871.448 &0.235 &V &+306 &12.637 &8 &\citet{de77} \\
2442871.503 &0.137 &B &+306 &12.693 &8 &\citet{de77} \\
2444012.939 &1.096 &V &+344 &15.170 &5 &\citet{ha80} \\
2444285.002 &0.045 &V &+353 &17.480 &12 &\citet{eg83} \\
2444465.335 &0.027 &V &+359 &17.976 &13 &\citet{cc85} \\
2444465.395 &0.023 &B &+359 &18.036 &12 &\citet{cc85} \\
2444525.621 &0.041 &B &+361 &18.317 &33 &\citet{eg83} \\
2444645.837 &0.051 &V &+365 &18.643 &21 &\citet{eg83} \\
2444946.666 &0.021 &B &+375 &19.746 &10 &\citet{cc85} \\
2444946.681 &0.020 &V &+375 &19.761 &10 &\citet{cc85} \\
2445006.453 &0.038 &V &+377 &19.587 &29 &\citet{mb84} \\
2445006.810 &0.018 &B &+377 &19.944 &28 &\citet{mb84} \\
2445667.740 &0.194 &V &+399 &21.477 &7 &\citet{be86} \\
2445668.115 &0.122 &B &+399 &21.852 &6 &\citet{be86} \\
2448106.960 &0.021 &V &+480 &32.916 &79 &\citet{esa97} \\
2448558.731 &0.028 &V &+495 &35.098 &75 &\citet{esa97} \\
2448859.950 &0.032 &V &+505 &36.590 &27 &\citet{esa97} \\
2449793.836 &0.020 &V &+536 &41.325 &25 &\citet{bt95} \\
2449793.855 &0.014 &B &+536 &41.344 &24 &\citet{bt95} \\
2450095.059 &0.028 &B &+546 &42.822 &10 &\citet{ber02} \\
2450095.182 &0.033 &V &+546 &42.946 &10 &\citet{ber02} \\
2450456.767 &0.020 &B &+558 &44.859 &18 &\citet{ber02} \\
2450456.785 &0.010 &V &+558 &44.877 &17 &\citet{ber02} \\
2450547.137 &0.011 &V &+561 &45.311 &20 &\citet{bt98} \\
2450878.613 &0.015 &V &+572 &47.088 &33 &\citet{bt00} \\
2451240.167 &0.253 &V &+584 &48.971 &35 &\citet{bt01a} \\
2451241.151 &0.279 &B &+584 &49.955 &35 &\citet{bt01a} \\
2451631.535 &0.341 &V &+597 &50.695 &23 &\citet{bc01} \\
2451933.341 &0.013 &V &+607 &52.776 &66 &\citet{po02} \\
2451933.364 &0.031 &V &+607 &52.798 &17 &\citet{bt01b} \\
2451933.385 &0.022 &B &+607 &52.819 &17 &\citet{bt01b}  \\
2452174.563 &0.031 &V &+615 &54.216 &36 &\citet{po02} \\
2452325.317 &0.016 &V &+620 &55.107 &38 &\citet{bt04a}  \\
2452325.355 &0.018 &B &+620 &55.145 &37 &\citet{bt04a}  \\
2452656.932 &0.013 &V &+631 &57.024 &42 &\citet{bt04b}  \\
2452687.050 &0.011 &V &+632 &57.169 &83 &\citet{po02} \\
2452988.481 &0.024 &V &+642 &58.874 &32 &\citet{bt04c}  \\
2453048.828 &0.010 &V &+644 &59.276 &86 &\citet{po02} \\
2453440.655 &0.016 &V &+657 &61.458 &64 &\citet{po02} \\
2453772.192 &0.013 &V &+668 &63.297 &97 &\citet{po02} \\
2454164.248 &0.019 &V &+681 &65.710 &61 &\citet{po02} \\
2454495.959 &0.017 &V &+692 &67.722 &107 &\citet{po02} \\
2454888.110 &0.016 &V &+705 &70.228 &98 &\citet{po02} \\
\enddata
\end{deluxetable}

A potential complication is an unknown rate of mass loss for AQ Pup. \citet{ne12} have noted that enhanced mass loss in Cepheids can elevate the rate of period increase for Cepheids displaying positive period changes, implying that a rapid rate of mass loss might explain the high rate of period increase for AQ Pup relative to other Cepheids of similar period in the third crossing of the instability strip. The Cepheid has an atmospheric composition, [Fe/H] $= -0.08\pm0.01$ \citep{ro08}, indicating slightly less than solar metallicity. A more detailed spectroscopic study centered on abundances of the CNO elements, to test for dredge-up material for example, is not necessarily useful for establishing evolutionary status \citep{tb04} because of possible meridional mixing during the main-sequence stage \citep{ma01}. A search for features indicative of mass loss would be more informative.

\begin{figure}[!t]
\epsscale{0.90}
\plotone{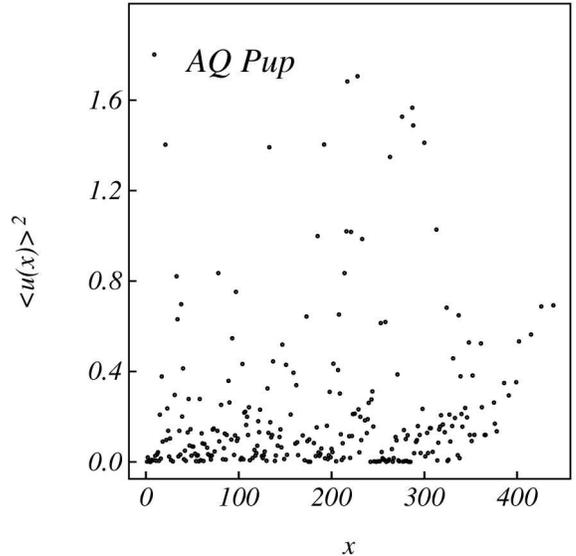}
\caption{\small{An Eddington-Plakidis test of the residuals from the observed times of light maximum for AQ Pup displays no evidence for random fluctuations in period. If that were the case, the data would scatter about a positive slope rather than the null slope as indicated.}}
\label{fig12}
\end{figure}

The residuals from the parabolic fit to the O--C data are plotted in the lower part of Fig.~\ref{fig10}. They do not appear to be randomly distributed, displaying instead a sinusoidal trend suggesting the possibility of light travel time effects in a binary system. Unfortunately, the implied orbital period of $\sim 84$ years and semi-major axis of $\sim1.2$ light days produce an uncomfortably large minimum total mass for the co-orbiting stars, so such a possibility must be tested by other means, from radial velocity measures for example. An Eddington-Plakidis test \citep{ep29} on the residuals shown in Fig.~\ref{fig12} also displays no evidence for random fluctuations in period \citep[see][]{te09b}. Chaotic period fluctuations in AQ Pup would normally be revealed by an upwards slope to the plotted data. Unless the tabulated individual times of light maximum are affected by temporal averaging \citep[see][]{te09b}, there is no evidence to suggest that the residuals of Fig.~\ref{fig10} arise from random fluctuations in period. Perhaps they are indicative of episodic variations in the rate of mass loss in AQ Pup, which additional observations could test.

\begin{deluxetable}{@{\extracolsep{-2.5mm}}lcc}
\tabletypesize{\small}
\tablewidth{0pt}
\tablenum{4}
\tablecaption{Derived parameters for AQ Pup and V620 Pup
\label{tab4}}
\tablehead{
\colhead{Parameter} &\colhead{AQ Pup} &\colhead{V620 Pup} }
\startdata
log P &1.4767 &0.4126  \\
log \.{P} (s yr$^{-1}$) &$2.4772\pm0.0017$ & ... \\
$E_{B-V}$ &$0.47\pm0.07$ &$0.64\pm0.02$ \\
($\langle B \rangle - \langle V \rangle$)$_0$ &+0.89 &+0.52 \\
$\langle M_V \rangle$ &$-5.40\pm0.25$ &$-2.74\pm0.11$ \\
$\log t$ (yr) &7.5 &8.0 \\
$d$ (pc) &$3213\pm185$ &$2880\pm107$ \\
\enddata
\end{deluxetable}

The inferred parameters for the two Cepheids are summarized in Table~\ref{tab4}. The derived space reddening for AQ Pup compares well with a {\it BVI} reddening of $E_{B-V}=0.504$ derived by \citet{lc07} and a model atmosphere reddening of $E_{B-V}=0.453$ found by \citet{ko08}. The inferred distance of $3.21\pm0.19$ kpc to AQ Pup found here also agrees exactly with the estimate of 3.21 kpc derived by \citet{st11} using the infrared surface brightness version of the Baade-Wesselink method. However, the values of $E_{B-V}=0.518$ and $\langle M_V \rangle = -5.51$ inferred for AQ Pup by \citet{st11} differ slightly from the present results. The implied luminosity of AQ Pup as a member of Turner~14 is $\langle M_V \rangle = -5.40\pm0.25$ with the uncertainty in reddening included. The difference relative to the \citet{st11} results is small, and the parameters are consistent with a Cepheid lying near the center of the instability strip, as implied by its relatively large light amplitude of $\Delta B = 1.84$. The agreement might be optimized if AQ Pup were in a putative fifth crossing lying slightly towards the cool edge of the instability strip. The derived luminosity of AQ Pup is otherwise exactly that predicted by the period-radius and color-effective temperature relations of \citet{tb02}, providing further confirmation of their validity.

The photometric parallax of $\pi = 0.31\pm0.02$ mas derived here for AQ Pup can be compared with parallaxes from the Hipparcos mission of $8.85\pm4.03$ mas \citep{esa97} and $15.47\pm3.59$ mas \citep{vl07}, a difference of more than $2\sigma$ for the original Hipparcos estimate and more than $4\sigma$ for the revised value. Conceivably the discrepancy arises from contamination by the nearby companions to AQ Pup, as suggested by \citet{sz97} for other Cepheids observed by Hipparcos.

V620 Pup is not studied well enough to provide a comparison with previous studies. The implied space reddening yields an intrinsic $(\langle B \rangle-\langle V \rangle)_0$ color of +0.52 and a luminosity as a member of Turner~13 of $\langle M_V \rangle = -2.74\pm0.11$. The luminosity is $\sim0.40$ magnitude more luminous than would be predicted by the relations of \citet{tb02} for a 2$^{\rm d}$.59 classical Cepheid, but that could be explained if the Cepheid lies on the blue side of the instability strip (as for the case of a first crossing) or is an overtone pulsator. The derived intrinsic color appears to indicate a Cepheid lying near strip center, or blueward of strip center for the case of overtone pulsation. The implied isochrone fit for Turner~13 in Fig.~\ref{fig8} indicates that V620 Pup is in the first crossing, which, if true, would be accompanied by measurable increases in pulsation period over a short time interval. The Cepheid has not been observed long enough to test such a possibility, but archival images may contain information for investigating that further.

A twelfth magnitude star $\sim0\arcmin.5$ southeast of V620 Pup appears to be a reddened F-type star, possibly a giant. It is the object 0.86 magnitude more luminous than the Cepheid in Fig.~\ref{fig8}. The photometric analysis suggests that it may also lie close to the Cepheid instability strip, provided it is an evolved cluster member. However, there is no indication of variability in the star from photometric monitoring of the V620 Pup field.

\section{{\rm \footnotesize CONCLUSIONS}}
Although there are no obvious OB associations in the region, a {\it UBV} survey of a field located around the Cepheids AQ Pup and V620 Pup reveals the presence of two putative clusters: one centered near V620 Pup that appears to contain the Cepheid as a likely member, and one centered $\sim 17\arcmin$ south of AQ Pup, with outlying stars surrounding the Cepheid, that appears to contain the $29^{\rm d}.97$ pulsator as a member. Both clusters are faint, poorly-populated, and near the limits of imaging surveys like the Palomar and ESO-SRC atlases. Their existence is argued by the dereddened parameters of likely cluster members, including the Cepheids.

The possible association of AQ Pup with the cluster Ruprecht 43 suggested by \citet{ts66} has never been investigated fully, primarily because the coordinates cited for Ruprecht~43 appear to be in error. There is a star chain centered on J2000.0 coordinates 07:58:46, --28:48:47, that has the appearance of a small, faint, compact cluster, and the declination differs by only $10^{\prime}$ from the value cited for Ruprecht~43 by \citet{al12}. Possibly there was an error in the original coordinates cited for the cluster? There is also an overlooked faint cluster of stars located closer to AQ Pup at J2000.0 coordinates 07:59:19, --28:58:00, designated here as Turner-Majaess~1. Ruprecht~43 and Turner-Majaess~1 both lie outside of the area surveyed in Fig.~\ref{fig1}.

\begin{figure}[!t]
\epsscale{0.90}
\plotone{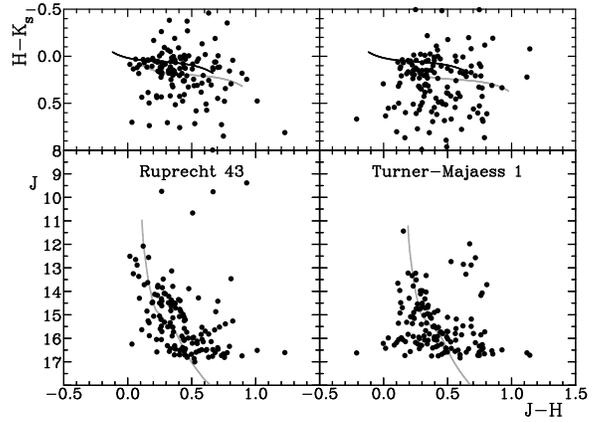}
\caption{\small{2MASS color-color (upper) and color-magnitude (lower) diagrams for measured stars in Ruprecht 43 (left) and Turner-Majaess~1 (right). The intrinsic color-color relation is plotted as a black curve, while gray curves represent the same relation for reddenings of $E_{B-V}=0.68$ (left) and $E_{B-V}=0.92$ (right). The plotted ZAMS relations (gray curves, lower) correspond to the same reddenings for the intrinsic distance modulus of Fig.\ref{fig7}.}}
\label{fig13}
\end{figure}

2MASS color-color and color-magnitude diagrams for stars lying within $2\arcmin$ of the adopted centers for the two clusters are presented in Fig.~\ref{fig13}. The deduced reddenings for the two clusters, $E_{B-V}=0.68$ and 0.92, respectively, were inferred using the techniques outlined by \citet{tu11}, while the distance moduli were chosen to yield intrinsic values identical to the best-fitting results of Fig.~\ref{fig7}, which apply to AQ Pup and Turner~14. They were not derived using best-fitting procedures, although it is noteworthy that the data for both clusters are a reasonably good match to the adopted values. Such good agreement argues that Ruprecht 43 and Turner-Majaess~1 probably lie at roughly the same distance as AQ Pup and Turner~14, suggesting that they belong to the same star complex. Deeper photometry is required to explore that conclusion further.

Absolute magnitudes have been derived for AQ Pup and V620 Pup under the assumption that they are members of the clusters in their vicinity. The case for AQ Pup seems reasonably strong, and the Cepheid appears to be in the third or possibly fifth crossing of the instability strip. Its rapid rate of period increase might also be indicative of rapid mass loss. V620 Pup is curious, given that the cluster match implies it is in the first crossing of the instability strip. But it has completely different characteristics from other putative first-crossers \citep{tu09}, with a very skewed light curve and a sizeable light amplitude of $\Delta B = 0.77$, more like that of second or third crossers. Further study of the Cepheid and the cluster surrounding it appears to be essential.

\subsection*{{\rm \scriptsize ACKNOWLEDGEMENTS}}
\scriptsize{David Pass and Mark Starzomski made star counts for the field of AQ Pup as part of a research experience project while students at Prince Edward High School in Dartmouth, Nova Scotia. The present study was supported by the National Research Council of Canada and by research funding awarded through the Natural Sciences and Engineering Research Council of Canada (NSERC), and the Russian Foundation for Basic Research (RFBR). We are indebted to the director of Harvard College Observatory for access to the plate stacks, and to the referee for useful suggestions.}

\end{document}